\documentclass[graybox]{svmult}

\usepackage{type1cm}        
\usepackage{makeidx}         
\usepackage{graphicx}        
\usepackage{multicol}        
\usepackage[bottom]{footmisc}

\usepackage{newtxtext}       %
\usepackage{newtxmath}       

\usepackage{psfrag}

\usepackage[colon,numbers,sort&compress]{natbib}
\makeindex             

\begin{document}

\title*{Configurational Forces in Penetration Processes}
\author{Davide Bigoni, Marco Amato and Francesco Dal Corso}
\authorrunning{D. Bigoni et al.}
\institute{D. Bigoni, M. Amato and F. Dal Corso \at
Department of Civil, Environmental and Mechanical Engineering,
University of Trento \\
Via Mesiano 77, I-38123 Trento, Italy \\
\email{bigoni@ing.unitn.it}
}
%
%
\maketitle
%
%

\vspace*{0.5cm}

\abstract{
With a loose reference to problems of penetration in biomechanics (for instance, a nanoparticle penetrating through a cell's membrane or a cell sucked with a pipette), the role of configurational forces is investigated during the process in which a compliant intruder is inserted into an elastic structure. For insertion into a rigid constraint, a configurational force proportional to the square of the strain needed to deform the body, which is penetrating, is found. This force has a more complex structure when the compliance of the constraint is kept into account, but in all cases, it tends to expel the penetrating body.
}

\section{Introduction}
\label{Introduction}
Biomechanics is a fast developing multidisciplinary science aimed at challenging the secrets of nature. The biomechanics of soft tissues, including successfully modeling of muscles (\citet{holz1}), blood vessels (\citet{holz2, holz3}), brain (\citet{holz4}; \citet{holz5}), and vascular tissues (\citet{holz6}), is recently oriented to a deeper understanding of the mechanics of the cell (\citet{disher2}; \citet{boal}; \citet{deseri2}; \citet{deseri3}; \citet{deseri1}; \citet{menzel}), a field strongly related to several timely problems, including the COVID-19 emergency. 

Membrane penetration of nano- or microparticles is a common feature in cell mechanics, important from several points of view, including drug delivery and viral entry. 
In particular, viral entry into a cell during the early stage of infection may occur in different forms. In the case of penetration, the cell's membrane is punctured after attachment of the virus, which in this way injects its contents inside the cytoplasm. 

Cells may be subject to large deformations, a situation occurring, for instance, when a cell is drawn up a pipette, as is the case of a human red blood cell (\citet{disher}). This process involves a competition between forces tending to inject and eject the cell. Similarly, but at a smaller scale, the influenza A virus can be trapped into a nanotube with an internal diameter of 400\,nm (\citet{yuge}). Membrane penetration and pipette suction are just two examples of problems involving the insertion of a compliant body inside a soft (the membrane) or a rigid (the pipette) structure. 
Many other problems of this kind can be listed in biomechanics: endoscopy, insertion of a catheter into a blood vessel or into the urinary tract, or injection of a needle for biopsy or for puncturing tissues (\citet{holz7}). 

\begin{figure}[t]
\begin{center}
\psfrag{a}[c][c]{\small (a)}
\psfrag{b}[c][c]{\small (b)}
\psfrag{c}[c][c]{\small (c)}
\psfrag{d}[c][c]{\small (d)}
\psfrag{e}[c][c]{\small (e)}
%
		\includegraphics[width=0.975\textwidth]{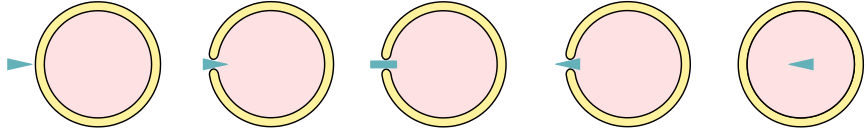}
%
%
\caption{Schematics of the penetration process: (a)~trapping, (b)~puncturing, (c)~opening, (d)~penetration, and (e)~end of the process. The penetrating particle and the cell are assumed to be deformable in a two-dimensional formulation, where friction, dissipation, and interparticle non-mechanical interactions are neglected.}
\label{penetrazione}       
\end{center}
\end{figure}

In the present article, the penetration of a compliant body into an elastic structure (such as a cell's membrane) or into a rigid constraint (as when a cell is drawn up a pipette) is analyzed from a purely mechanical perspective, under simple assumptions. These include 
restriction to the two-dimensional formulation (a gross, but common, approximation (\citet{menzel})) and absence of: (i)~friction, (ii)~dissipative forces, and (iii)~interparticle non-mechanical interactions, such as, for instance, magnetic attraction. 
The simple mechanical process of penetration schematized in Fig.~\ref{penetrazione} is addressed, with the specific purpose to explore and present the role of configurational forces, actions that develop when an elastic system can change its configuration through a release of elastic energy.
%
These forces have been introduced by \citet{eshelby} to describe the mechanics of defects and only recently 
explored in elastic structures subject to bending (\citet{conf1}) and torsion (\citet{conf2}). Configurational forces in structures lead to unexpected effects, such as possibility of 
self-encapsulation (\citet{conf3}) and expulsion during a penetration process (\cite{conf4}) due to bending flexibility of inextensible rods. 

\section{Penetration and Configurational Forces}
\label{Penetration and configurational forces}
A simple model is introduced, which may capture essential features of the \lq puncturing' and \lq penetrating' stages of the intrusion process, shown in Fig.~\ref{penetrazione}. 
Besides its application to mechanobiology, the model is formulated to introduce the role of the configurational force in the processes of insertion of an elastic body into another. 

\begin{figure}[b]
\begin{center}
\psfrag{a}[c][c]{\small (a)}
\psfrag{b}[c][c]{\small (b)}
\psfrag{k}[c][c]{\small $k$}
\psfrag{x}[c][c]{\small $x_1$}
\psfrag{y}[c][c]{\small $x_2$}
\psfrag{z}[c][c]{\small $x_3$}
\psfrag{i}[c][c]{\small $\bar{h}$}
\psfrag{f}[c][c]{\small $b_0$}
\psfrag{g}[c][c]{\small $h_0$}
\psfrag{c}[c][c]{\small $b$}
\psfrag{h}[c][c]{\small $h$}
\psfrag{l}[c][c]{\small $l$}
%
				\includegraphics[width=.975\textwidth]{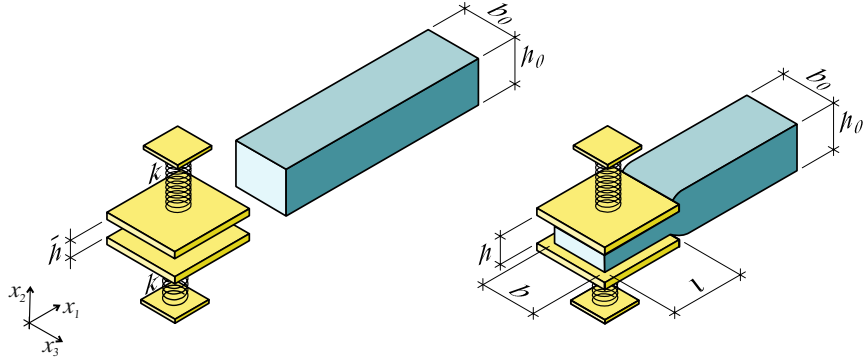}
%
%
\caption{An elastic layer is inserted between two rigid, flat, and frictionless plates (remaining parallel to the $x_1-x_3$ plane), each  held by a linear spring of elastic stiffness $k$. The distance between the two rigid plates at rest is $\bar{h}$, smaller than the height $h_0$ of the elastic layer (a prism of initial cross-section $h_0 \times b_0$) in its unstressed configuration. In the deformed configuration, the prism height is $h\in \left[\bar{h},h_0\right]$, the width is $b\geq b_0$, and the insertion length is $l$.}
\label{modellino}       
\end{center}
\end{figure}

A deformable elastic layer is considered of initial height $h_0$, out-of-plane width $b_0$, and length for the moment left unspecified. This layer has to be inserted between
two rigid planar and frictionless constraints, placed at a distance $\bar{h} \leq h_0$ along the $x_2$-axis when the two linear elastic springs of stiffness $k$ are unloaded, see Fig.~\ref{modellino}(a). These two rigid plates are prescribed to remain parallel to the $x_1-x_3$ plane.

The spring stiffness $k$ simply models the elastic stiffness of the structure in which the intruder has to penetrate. For instance, when the structure is a thin elastic ring 
(as sketched in Fig.~\ref{penetrazione}) of radius $R$ and bending stiffness $EJ$, the spring stiffness becomes $k=2EJ/(3\pi R^3)$. 
The elastic layer is partially inserted (of an amount $l>0$) between the two frictionless constraints so that a deformed configuration such as that illustrated in Fig.~\ref{modellino}(b) is realized, where the  height of the inserted layer is $h \in  \left[\bar{h},h_0\right]$. 
Although near the edge of the constraint exit, the stress/strain state is highly disuniform, the unloaded configuration is reached at a sufficient distance
far from the constraint's exit, while a state of uniaxial compression is approached inside the rigid plates (where the width of the layer becomes $b \geq b_0$). 
Therefore, the edges of the constraint induce a strong disturbance, but the stress/strain state tends to become uniform 
far from this point, when the elastic layer is sufficiently long. 
Since the contact between the layer and the constraint is smooth, one would be tempted to conclude that no horizontal forces are applied to the elastic body along the $x_1$-axis. This too facile conclusion is missing the forces which develop at the corners of the constraint, representing the \lq microscopic' counterpart of the configurational force concept. 
In particular, the configurational force can be understood as follows. The total potential energy $\mathcal{V}$  of the deformed system represented in Fig.~\ref{modellino} is coincident with the elastic energy $\mathcal{E}$, which is given by the sum of that stored in the springs and in the layer. This energy has to depend on the configurational parameter $l$, i.e. the amount of layer inside the constraint. If $l$ decreases, the energy $\mathcal{V}(l)=\mathcal{E}(l)$ will decrease too, until becoming null when the layer loses contact with the rigid plates because outside ($l\leq 0$). \citet{eshelby}  defined the configurational force $P$ acting on the elastic system as the negative of the partial derivative of the total potential energy $\mathcal{V}(l)$ at equilibrium with respect to the configurational parameter. In the present case, the configurational force reduces to
\begin{equation}
\label{eq:hyperelasticity}
P=-\frac{\mbox{d} \mathcal{E}(l)}{\mbox{d} l},
\end{equation}
where the \lq $-$' sign arises from the fact that on increasing $l$, the layer moves in the negative direction of the $x_1$-axis. 
 Therefore, a configurational force emerges, which pushes the layer outside the constraint and can be viewed from a \lq microscopic' point of view as the tangential reaction at the frictionless sliding constraint, provided as the resultant of the actions realized at the constraints corners. 

The elastic energies stored in the layer and in the springs can be evaluated in an approximate way as follows. 
The strain-energy density function $W$ for an incompressible Mooney-Rivlin material is (\citet{bigoni_libro})
\begin{equation}
W(\lambda_1, \lambda_2, \lambda_3) = \frac{\mu_1}{2} \left(\lambda_1^2 + \lambda_2^2 + \lambda_3^2 -3\right) -
\frac{\mu_2}{2} \left(\frac{1}{\lambda_1^{2}} + \frac{1}{\lambda_2^{2}} +\frac{1}{\lambda_3^{2}} -3\right),
\end{equation}
where $\mu_1$ and $\mu_2$ are material constants, while $\lambda_1$, $\lambda_2$, and $\lambda_3$ are the principal stretches, the latter subject to the incompressibility constraint
\begin{equation}\label{incomp}
\lambda_1 \lambda_2 \lambda_3 = 1.
\end{equation}
The moduli $\mu_1$ and $\mu_2$ are subject to the restrictions
\begin{equation}\label{PD}
\mu_1 \geq 0, \qquad \mu_2 \leq 0,
\end{equation}
and values representative for the behavior of rubber at room temperature are $\mu_1 = 3$\,bar and $\mu_2 = -0.3$\,bar. 

Consider a parallelepiped of initial height $h_0$ and transverse dimensions $l_0$ and $b_0$, compressed parallel to the edge $h_0$ with a uniaxial state of stress, reducing the height to $h$ and enlarging the other two dimensions to $l$ and $b$. 
The stretches are
\begin{equation}
\lambda_1 = \frac{l}{l_0},  \qquad \lambda_2 = \frac{h}{h_0}, \qquad \lambda_3 = \frac{b}{b_0},  
\end{equation}
which, due to loading symmetry ($\lambda_1=\lambda_3$) and material incompressibility (eq.~\eqref{incomp}), are constrained by
\begin{equation}
\lambda_1 = \frac{l}{l_0} = \lambda_3 = \frac{b}{b_0} = \sqrt{\frac{h_0}{h}}.   
\end{equation}
Therefore, the elastic energy stored in the uniformly deformed parallelepiped corresponds to 
\begin{equation}
\label{palla}
\begin{array}{ll}
l_0 b_0 h_0  \, W =  \displaystyle l \sqrt{\frac{h}{h_0}} \, b_0 h_0 \, \left[
\frac{\mu_1}{2} \left(\frac{h^2}{h_0^2} + 2 \frac{h_0}{h}-3\right) 
-
\frac{\mu_2}{2} \left(\frac{h_0^2}{h^2} + 2\frac{h}{h_0} -3\right) 
\right],
\end{array}
\end{equation}
where the dependency on the parameter $l$, crucial in the following calculations, has been evidenced. 

When the stretch $\lambda_2=h/h_0$ is close to 1, the energy (eq.~\eqref{palla}) can be approximated by
\begin{equation}
\label{piccolina}
l_0 b_0 h_0  \, W = \frac{3}{2} (\mu_1 - \mu_2)\,l\,b_0\,h_0 \epsilon^2, 
\end{equation}
where $\epsilon$ is the infinitesimal strain along the $x_2$-axis
\begin{equation}
\epsilon = \dfrac{h}{h_0}-1,
\end{equation}
and for small strain, the elastic energy (eq.~\eqref{palla}) reduces for $\mu_1=\mu$ and $\mu_2=0$ to the  strain energy of an isotropic  and incompressible linear elastic solid.

The energy, as given in eq.~\eqref{palla}, provides a simple approximation to that stored in the whole deformed elastic layer. The contribution of the highly inhomogeneous zone near the 
edge of the constraint is completely neglected, but this approximation becomes reasonable when the parts of the layer inside the constraint and outside are sufficiently long. In this way, the variation of the stored energy when a part of the layer is expelled from the constraint corresponds with a good approximation to the final segment of the layer inside the constraint, which is subject to a stress/strain state approximately uniform and corresponding to uniaxial stress.

\subsection{Penetration of a Rigid Body} 
\label{Rigid channel}

If the penetrated body in which the elastic intruder is inserted is rigid, $k\rightarrow\infty$, the height $h=\bar{h}$ is fixed, and no energy is stored within the springs. In these conditions and for the above considerations, the configurational force $P$ can be easily obtained through differentiation of eq.~\eqref{palla} with respect to the parameter $l$, i.e.
\begin{equation}
\label{palladio}
\frac{P}{\mu_1b_0h_0} = - \frac{1}{2} \, \sqrt{\frac{h}{h_0}} \, \left[
\frac{h^2}{h_0^2} + 2 \frac{h_0}{h}-3 -
\frac{\mu_2}{\mu_1} \left(\frac{h_0^2}{h^2} + 2\frac{h}{h_0} -3\right) 
\right].
\end{equation}
Using the conditions stated in eq.~\eqref{PD}, it can be seen from eq.~\eqref{palladio} that the configurational force is negative ($P<0$), and therefore, it tends to expel the layer from the constraint (in other words, to move it in the positive direction of the $x_1$-axis). 

Equation~\eqref{palladio} is a complex function of the elastic energy via the stretch $\lambda_2=h/h_0$ and shows the independence of the length $l$ for the configurational force $P$. 
When the strain $\epsilon$ is small, the configurational force $P$  can be approximated as a quadratic expression of the infinitesimal strain $\epsilon$, i.e.
\begin{equation}
\label{palladio2}
\frac{P}{\mu_1b_0h_0} = - \frac{3}{2} \left(1 - \frac{\mu_2}{\mu_1}\right) \epsilon^2.
\end{equation}

\subsection{Role of Penetrated Body's Elasticity}
\begin{figure}[t]
\begin{center}
				\includegraphics[width=.9\textwidth]{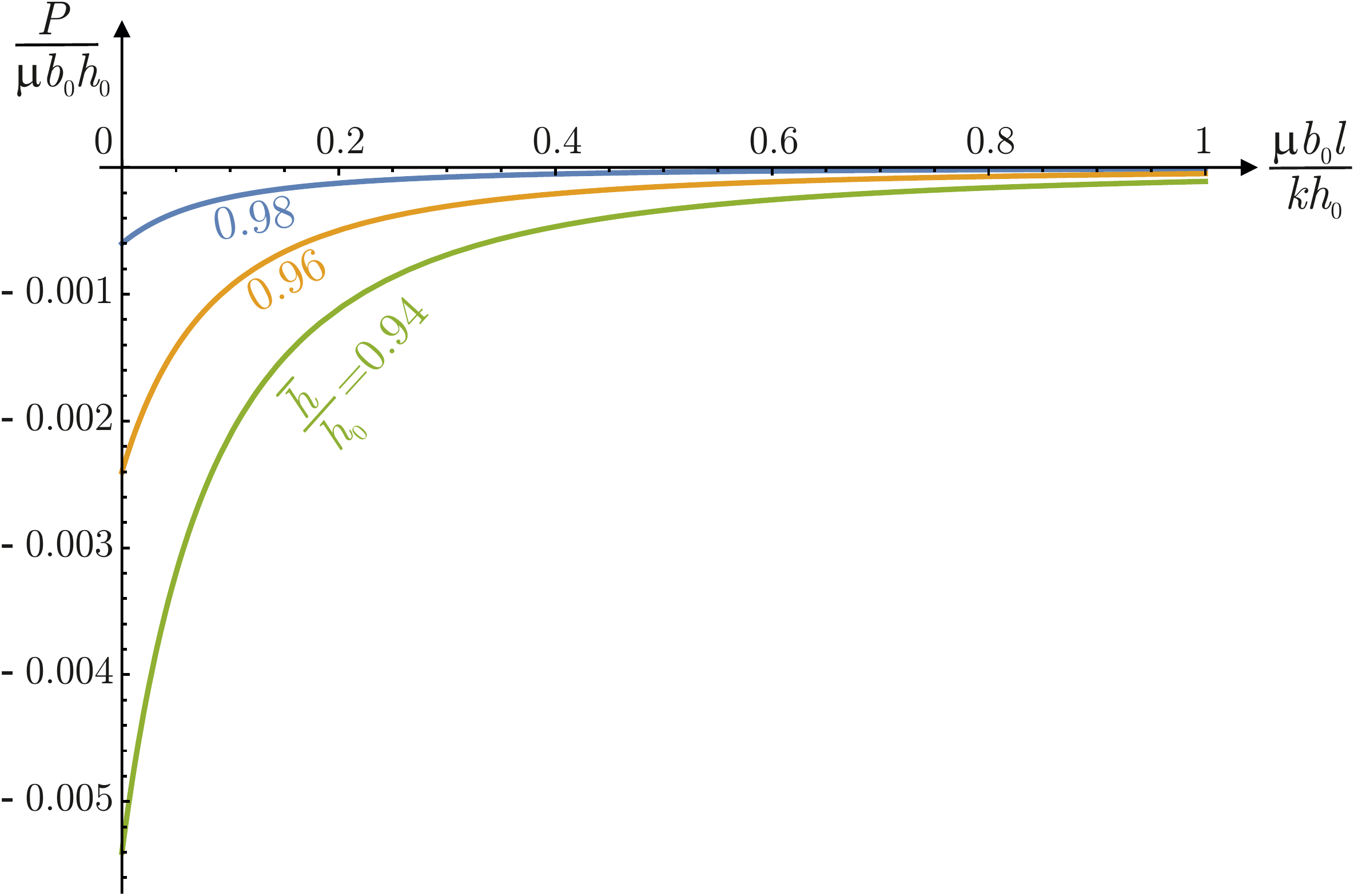}
%
%
\caption{Configurational force $P$, made dimensionless through division by $\mu b_0 h_0$, as a function of $\mu b_0 l /(k h_0)$ for three values of the ratio $\bar{h}/h_0=\{0.94,0.96,0.98\}$.}
\label{figurata}       
\end{center}
\end{figure}

In this section, the effect of the elasticity of the structure penetrated by the intruding layer is accounted for by considering each rigid plate suspended by a linear elastic spring of stiffness $k$, as illustrated in Fig.~\ref{modellino}. 
The total elastic energy $\mathcal{E}$, consisting of the energies stored in the two springs of stiffness $k$ and the elastic layer, is
\begin{equation}
\label{energiazza}
\begin{array}{ll}
\displaystyle \frac{\mathcal{E}}{\mu_1b_0h_0}  = \displaystyle \frac{k h_0}{4\mu_1b_0} \left( \frac{h}{h_0}- \frac{\bar{h}}{h_0}\right)^2 +
 \displaystyle \frac{l}{2} \sqrt{\frac{h}{h_0}} \left[
\frac{h^2}{h_0^2} + 2 \frac{h_0}{h}-3 
-
\frac{\mu_2}{\mu_1} \left(\frac{h_0^2}{h^2} + 2\frac{h}{h_0} -3\right) 
\right].
\end{array}
\end{equation}
In this case, the height $h$ is not imposed, as in eq.~\eqref{palladio}, but is an unknown, which depends on the relative stiffness 
between the springs and the elastic layer. 
To proceed further, the small strain assumption is introduced, together with $\mu_1=\mu$ and $\mu_2=0$.
Under these restrictions, the configurational force can be easily calculated as follows. 
The total elastic energy $\mathcal{E}$ (eq.~\eqref{energiazza}) becomes a function of the (small) strain $\epsilon$ in the layer and reads
\begin{equation}
\label{energiazza2} 
\frac{\mathcal{E} (\epsilon,l)}{\mu b_0h_0} =  \frac{k h_0}{4\mu b_0} \left( 1+\epsilon - \frac{\bar{h}}{h_0}  \right)^2 +
\frac{3}{2} l\epsilon^2. 
\end{equation}
The stationarity of the total energy $\mathcal{E} (\epsilon,l)$, as introduced in eq.~\eqref{energiazza2}, with varying the strain $\epsilon$ in the layer gives the strain $\epsilon^\ast$ at equilibrium for a given length $l$, i.e.
\begin{equation}
\epsilon^\ast(l) = -\dfrac{1-\dfrac{\bar{h}}{h_0}}{ 1+\displaystyle \frac{6\mu b_0 l }{kh_0}} 
\end{equation}
(a negative quantity because of $\bar{h}/h_0<1$). Hence, the total elastic energy (eq.~\eqref{energiazza2}) evaluated at $\epsilon^\ast$ is
\begin{equation}
\label{plaf}
\frac{\mathcal{E} (\epsilon^\ast(l),l)}{\mu b_0h_0} =  \frac{k h_0}{4\mu b_0} 
\dfrac{\left( 1 - \dfrac{\bar{h}}{h_0}  \right)^2}{1+ \dfrac{k h_0}{6\mu b_0 l}}. 
\end{equation}
%
Using eq.~\eqref{eq:hyperelasticity}, the configurational force $P$ reads
\begin{equation}
\frac{P}{\mu b_0h_0}  = - 
\frac{3}{2} \left(\dfrac{1-\dfrac{\bar{h}}{h_0}}{ 1+\displaystyle \frac{6\mu b_0 l }{kh_0}}\right)^2
\end{equation}
and is reported in Fig.~\ref{figurata} 
as a function of $\mu b_0 l /(k h_0)$, for three values of the ratio $\bar{h}/h_0$, chosen close to one. At vanishing $\mu b_0 l /(kh_0)$, a finite limit is attained for the force $P$, i.e. 
\begin{equation}
\lim_{\frac{\mu b_0 l }{kh_0}\rightarrow 0}\frac{P}{\mu b_0h_0}  = - 
\frac{3}{2} \left(1-\dfrac{\bar{h}}{h_0}\right)^2,
\end{equation}
which corresponds to the value obtained under the approximation that the penetrated body is rigid (see Eq.~\eqref{palladio2}), by identifying the distance $\bar{h}$ between the rigid plates at rest with $h$.

\section{Conclusion}
Penetration of one body into another is a common process in biomechanics. It has been shown in the present note that configurational forces may play an important role in these processes. The configurational forces can be calculated by considering the stored elastic energy and represent the counterpart of the microscopic actions developing at the moving boundary points, present at the contact between the intruder and the penetrated body.

\paragraph{\textbf{Acknowledgments.}
Financial support is acknowledged from H2020-MSCA-ITN-\linebreak 2020-LIGHTEN-956547 and ARS01-01384-PROSCAN.}

\makeatletter
\renewcommand\@biblabel[1]{#1.}
\makeatother
\bibliography{jou-abb,references.bib}

\begin{thebibliography}{21}
\providecommand{\natexlab}[1]{#1}
\providecommand{\url}[1]{{#1}}
\providecommand{\urlprefix}{URL }
\expandafter\ifx\csname urlstyle\endcsname\relax
  \providecommand{\doi}[1]{DOI~\discretionary{}{}{}#1}\else
  \providecommand{\doi}{DOI~\discretionary{}{}{}\begingroup
  \urlstyle{rm}\Url}\fi
\providecommand{\eprint}[2][]{\url{#2}}

\bibitem[{Bigoni(2012)}]{bigoni_libro}
Bigoni, D.: Nonlinear Solid Mechanics: Bifurcation Theory and Material
  Instability.
\newblock Cambridge University Press (2012)

\bibitem[{Bigoni et~al.(2014{\natexlab{a}})Bigoni, Bosi, {Dal Corso}, and
  Misseroni}]{conf4}
Bigoni, D., Bosi, F., {Dal Corso}, F., Misseroni, D.: Instability of a
  penetrating blade. J. Mech. Phys. Solids \textbf{64}, 411--425 (2014)

\bibitem[{Bigoni et~al.(2014{\natexlab{b}})Bigoni, {Dal Corso}, Misseroni, and
  Bosi}]{conf2}
Bigoni, D., {Dal Corso}, F., Misseroni, D., Bosi, F.: Torsional locomotion.
  Proc. Math. Phys. Eng. Sci. \textbf{470}, 20140599 (2014)

\bibitem[{Bigoni et~al.(2015)Bigoni, {Dal Corso}, Bosi, and Misseroni}]{conf1}
Bigoni, D., {Dal Corso}, F., Bosi, F., Misseroni, D.: Eshelby-like forces
  acting on elastic structures: theoretical and experimental proof. Mech.
  Mater. \textbf{80}, 368--374 (2015)

\bibitem[{Boal(2002)}]{boal}
Boal, D.: Mechanics of the Cell.
\newblock Cambridge University Press, Cambridge (2002)

\bibitem[{Bosi et~al.(2015)Bosi, Misseroni, {Dal Corso}, and Bigoni}]{conf3}
Bosi, F., Misseroni, D., {Dal Corso}, F., Bigoni, D.: Self-encapsulation, or
  the \lq dripping' of an elastic rod. Proc. Math. Phys. Eng. Sci.
  \textbf{471}, 20150195 (2015)

\bibitem[{Daddi-Moussa-Ider et~al.(2019)Daddi-Moussa-Ider, Goh, Liebchen,
  Hoell, Mathijssen, Guzm{\'a}n-Lastra, Scholz, Menzel, and L{\"o}wen}]{menzel}
Daddi-Moussa-Ider, A., Goh, S., Liebchen, B., Hoell, C., Mathijssen, A.J.,
  Guzm{\'a}n-Lastra, F., Scholz, C., Menzel, A.M., L{\"o}wen, H.: Membrane
  penetration and trapping of an active particle. J. Chem. Phys. \textbf{150},
  064906 (2019)

\bibitem[{Deseri and Zurlo(2013)}]{deseri2}
Deseri, L., Zurlo, G.: The stretching elasticity of biomembranes determines
  their line tension and bending rigidity. Biomech. Model. Mechanobiol.
  \textbf{12}, 1233--1242 (2013)

\bibitem[{Deseri et~al.(2016)Deseri, Pollaci, Zingales, and Dayal}]{deseri1}
Deseri, L., Pollaci, P., Zingales, M., Dayal, K.: Fractional hereditariness of
  lipid membranes: Instabilities and linearized evolution. J. Mech. Behav.
  Biomed. Mater. \textbf{58}, 11--27 (2016)

\bibitem[{Discher et~al.(1994)Discher, Mohandas, and Evans}]{disher}
Discher, D., Mohandas, N., Evans, E.: Molecular maps of red cell deformation:
  hidden elasticity and in situ connectivity. Science \textbf{266}, 1032--1035
  (1994)

\bibitem[{Discher et~al.(1998)Discher, Boal, and Boey}]{disher2}
Discher, D.E., Boal, D.H., Boey, S.K.: Simulations of the erythrocyte
  cytoskeleton at large deformation. \mbox{II. M}icropipette aspiration.
  Biophys. J. \textbf{75}, 1584--1597 (1998)

\bibitem[{Eshelby(1951)}]{eshelby}
Eshelby, J.D.: The force on an elastic singularity. Philos. Trans. Royal Soc. A
  \textbf{244}, 87--112 (1951)

\bibitem[{Franceschini et~al.(2006)Franceschini, Bigoni, Regitnig, and
  Holzapfel}]{holz4}
Franceschini, G., Bigoni, D., Regitnig, P., Holzapfel, G.A.: Brain tissue
  deforms similarly to filled elastomers and follows consolidation theory. J.
  Mech. Phys. Solids \textbf{54}, 2592--2620 (2006)

\bibitem[{Holzapfel et~al.(2000)Holzapfel, Gasser, and Ogden}]{holz2}
Holzapfel, G.A., Gasser, T.C., Ogden, R.W.: A new constitutive framework for
  arterial wall mechanics and a comparative study of material models. J.
  Elasticity \textbf{61}, 1--48 (2000)

\bibitem[{Holzapfel et~al.(2004)Holzapfel, Sommer, and Regitnig}]{holz3}
Holzapfel, G.A., Sommer, G., Regitnig, P.: Anisotropic mechanical properties of
  tissue components in human atherosclerotic plaques. J. Biomech. Eng.
  \textbf{126}, 657--665 (2004)

\bibitem[{Mihai et~al.(2017)Mihai, Budday, Holzapfel, Kuhl, and
  Goriely}]{holz5}
Mihai, L.A., Budday, S., Holzapfel, G.A., Kuhl, E., Goriely, A.: A family of
  hyperelastic models for human brain tissue. J. Mech. Phys. Solids
  \textbf{106}, 60--79 (2017)

\bibitem[{Pandolfi and Holzapfel(2008)}]{holz6}
Pandolfi, A., Holzapfel, G.A.: Three-dimensional modeling and computational
  analysis of the human cornea considering distributed collagen fibril
  orientations. J. Biomech. Eng. \textbf{130}, 061006 (2008)

\bibitem[{Roxhed et~al.(2007)Roxhed, Gasser, Griss, Holzapfel, and
  Stemme}]{holz7}
Roxhed, N., Gasser, T.C., Griss, P., Holzapfel, G.A., Stemme, G.:
  Penetration-enhanced ultrasharp microneedles and prediction on skin
  interaction for efficient transdermal drug delivery. IEEE J.
  Microelectromech. Syst. \textbf{16}, 1429--1440 (2007)

\bibitem[{St{\aa}lhand et~al.(2008)St{\aa}lhand, Klarbring, and
  Holzapfel}]{holz1}
St{\aa}lhand, J., Klarbring, A., Holzapfel, G.A.: Smooth muscle contraction:
  mechanochemical formulation for homogeneous finite strains. Prog. Biophys.
  Mol. Biol. \textbf{96}, 465--481 (2008)

\bibitem[{Terzi et~al.(2015)Terzi, Dayal, Deseri, and Deserno}]{deseri3}
Terzi, M.M., Dayal, K., Deseri, L., Deserno, M.: Revisiting the link between
  lipid membrane elasticity and microscopic continuum models. Biophys. J.
  \textbf{108}, 87a--88a (2015)

\bibitem[{Yuge et~al.(2017)Yuge, Akiyama, Ishii, Namkoong, Yagi, Nakai, Adachi,
  and Komatsu}]{yuge}
Yuge, S., Akiyama, M., Ishii, M., Namkoong, H., Yagi, K., Nakai, Y., Adachi,
  R., Komatsu, T.: Glycoprotein nanotube traps influenza virus. Chem. Lett.
  \textbf{46}, 95--97 (2017)

\end{thebibliography}

\end{document}